\documentclass[aps,pra,twocolumn,tightenlines,groupedaddress,showpacs]{revtex4-1}
\usepackage{epsfig}
\usepackage{amsmath}
\usepackage{amsfonts}
\usepackage{color}
\usepackage{graphicx}
\usepackage{subfigure}

\def\bra#1{{\langle#1|}}
\def\ket#1{{|#1\rangle}}

\def\expect#1{{\langle#1\rangle}}

\def\proj{{\hat{\cal P}}}

\def\H{{\hat H}}

\def\a{{\hat a}}
\def\adag{{\hat a}^\dagger}

\def\bdag{{\hat b}^\dagger}

\def\A{\hat{A}}
\def\Adag{\hat{A}^\dagger}

\def\Sp{\hat{\sigma}^+}
\def\Sm{\hat{\sigma}^-}

\def\projX1{{\hat{\cal P}_{X_1}}}
\def\projX{{\hat{\cal P}_{X}}}

\def\probX1{\expect{\proj_{X_1}}}
\def\probX{\expect{\proj_{X}}}

\begin{document}
\title{Continuous monitoring can improve single-photon probability}
\author{ Shesha Raghunathan}
\email{sraghuna@usc.edu} 
\author{Todd Brun }
\email{tbrun@usc.edu}
\affiliation{Center for Quantum Information Science and Technology,
Communication Sciences Institute, Department of Electrical Engineering, 
University of Southern California, Los Angeles, CA 90089, USA.}

%
\begin{abstract}
An engineering technique using continuous quantum measurement together with a change detection algorithm is proposed to improve the probability of single photon emission for a quantum-dot based single-photon source. The technique involves continuous monitoring of the emitter, integrating the measured signal, and a simple change detection circuit to decide when to stop pumping. The idea is to pump just long enough such that the emitter $+$ cavity system is in a state that can emit at most one photon with high probability.  Continuous monitoring provides partial information on the state of the emitter. This technique is useful when the system is operating in the weak coupling regime, and the rate of pumping is smaller than, or comparable to, the emitter-cavity coupling strength, as can be the case for electrical pumping.
\end{abstract}

\maketitle

\section{Introduction}
Generation of single-photon states has wide-ranging applications, spanning quantum computing, quantum imaging, metrology, communication, and cryptography, amongst others~\cite{SPS-reviews}.  These applications are important to scientific and technological progress in many important areas. For the current work, we are especially interested in Linear Optical Quantum Computation (LOQC) \cite{Knill01,Kok07}. A key requirement of LOQC is the availability of high quality single-photon states on demand. 

Semiconductor quantum dot-based implementations of single-photon sources are of particular interest, as they scale well upon integration and are amenable to commercial fabrication techniques \cite{SPS-reviews, SPS-QD-implementations}. Typically, implementation of these devices involves a quantum dot (QD or dot) inside a microcavity, with the dot acting as the photon source:  the presence of the cavity increases the collection efficiency due to spatial confinement of the photons \cite{SPS-reviews, SPS-QD-implementations, Michler-Vahala03}. A wide variety of cavities with different sizes, shapes, and quality factors have been designed and fabricated \cite{Michler-Vahala03}. 

A quantum dot in a microcavity can be pumped either optically or electrically. Optical pumping is more straightforward experimentally, but electrical pumping may be better suited to large scale integration.  Also, as it does not directly insert photons into the cavity, it opens up the possibility of pumping directly into an energy level resonant with the cavity mode, which may reduce timing uncertainties in photon emission \cite{Raghunathan09}.  It also allows a channel for continuous measurement of the dot.  In electrical pumping, a bias voltage is applied across a quantum dot $p-n$ diode that enables an electron to tunnel through from $n$-type onto the dot (present at $p-n$ junction). Once an electron tunnels through to the dot, no new electron can tunnel through due to the Coulomb blockade effect \cite{electrical-pumping}. A further increase in the bias voltage then enables a hole from the $p$-type to tunnel to the dot. Recombination of electron and hole in the dot follows, leading to a photon emission into the cavity.  Recombination also leads to a drop in the potential across the quantum dot $p-n$ diode. By observing this change in the potential across the quantum dot $p-n$ diode, we can gain information about the state of the dot. This observation can be seen as a weak quantum measurement that provides partial information about the system \cite{Levi07}. 

A good single-photon source should be able to produce exactly one photon at the required time, in a specified state. {\it Indistinguishability} is a measure that captures the specificity of the photon state (due, e.g., to time-uncertainty of photon emission), while {\it single-photon probability} determines the likelihood that a single photon is indeed emitted. We have considered continuous monitoring as a tool to improve indistinguishability elsewhere~\cite{Raghunathan09}.  We focus on single-photon probability  here.

There are various processes that affect single-photon probability: collection efficiency (emitting into the cavity mode), pumping, cavity leakage to non-waveguide modes, and photon loss, to name a few. Many of these processes depend on the fabrication techniques and the materials that go with it, and hence can only be improved by building better sources.  Some, however, may be improved by better control.  The most obvious process to treat from this point of view is pumping:  how long should we pump at a given pumping strength to maximize the single-photon probability, given that other parameters are fixed?

In the case of strong pumping---as optical pumping often is---the answer is straight forward: pump for a duration short compared to the emission rate of the dot.  However, weakly pumped systems (like electrical pumping) present a more complicated scenario.  Since the pumping strength is weak, we might have to pump for times comparable to, or even longer than, the emission rate of the dot.  We thus may increase the likelihood of multi-photon emission. Since electrical pumping with semi-conductor QD based implementations is of active experimental interest \cite{electrical-pumping}, understanding the behavior of weakly pumped systems is critical.  

%
%
In this work, we use continuous quantum measurements to improve single-photon probability. The idea is simple:  to monitor the state of the emitter continuously, and determine when to stop pumping energy into the system based on the information obtained. Unfortunately, the output from monitoring such a microscopic system is intrinsically noisy, and little time is available to process the received signal.  This complicates the procedure.  We utilize a sequential statistical technique called CUmulative SUM (CUSUM) as the decision making process. We show numerically that this mechanism substantially improves single-photon probability in the weak coupling regime, particularly when the pumping rate is comparable to the QD-Cavity coupling strength.

\subsection{Overview of the paper}
The emitter we consider is a {\it p-n} diode operated as a single-photon LED, though the techniques we discuss are probably applicable to other systems.  We discuss the emitter in Sec.~\ref{sec:single-led}. We then capture the essential features of the LED's operation in a schematic model in Sec.~\ref{sec:system}; we present details regarding the system in a cavity-QED setting, and give its energy-level diagram; we present a stochastic master equation including the relevant physical processes, and discuss the parameter regime of operation. Section~\ref{sec:change-detection} presents the change detection algorithm and the decision circuit used to stop pumping; we consider sequential CUSUM technique in Sec.~\ref{sec:cusum} and present Bayesian approach in Sec.~\ref{sec:bayesian}. We give results in Sec.~\ref{sec:results}: we discuss the deterministic case first in Sec.~\ref{sec:det}, before proceeding to analyze the numerical performance of the CUSUM-based technique in Sec.~\ref{sec:perform}.  Finally, we scrutinize the effect of monitoring efficiency on performance in Sec.~\ref{sec:efficiency}. 

\section{Single-photon light emitting diode}
\label{sec:single-led}

\begin{figure}[htp]
  \begin{center}
    	 \includegraphics[width=2.5in,height=1.75in]{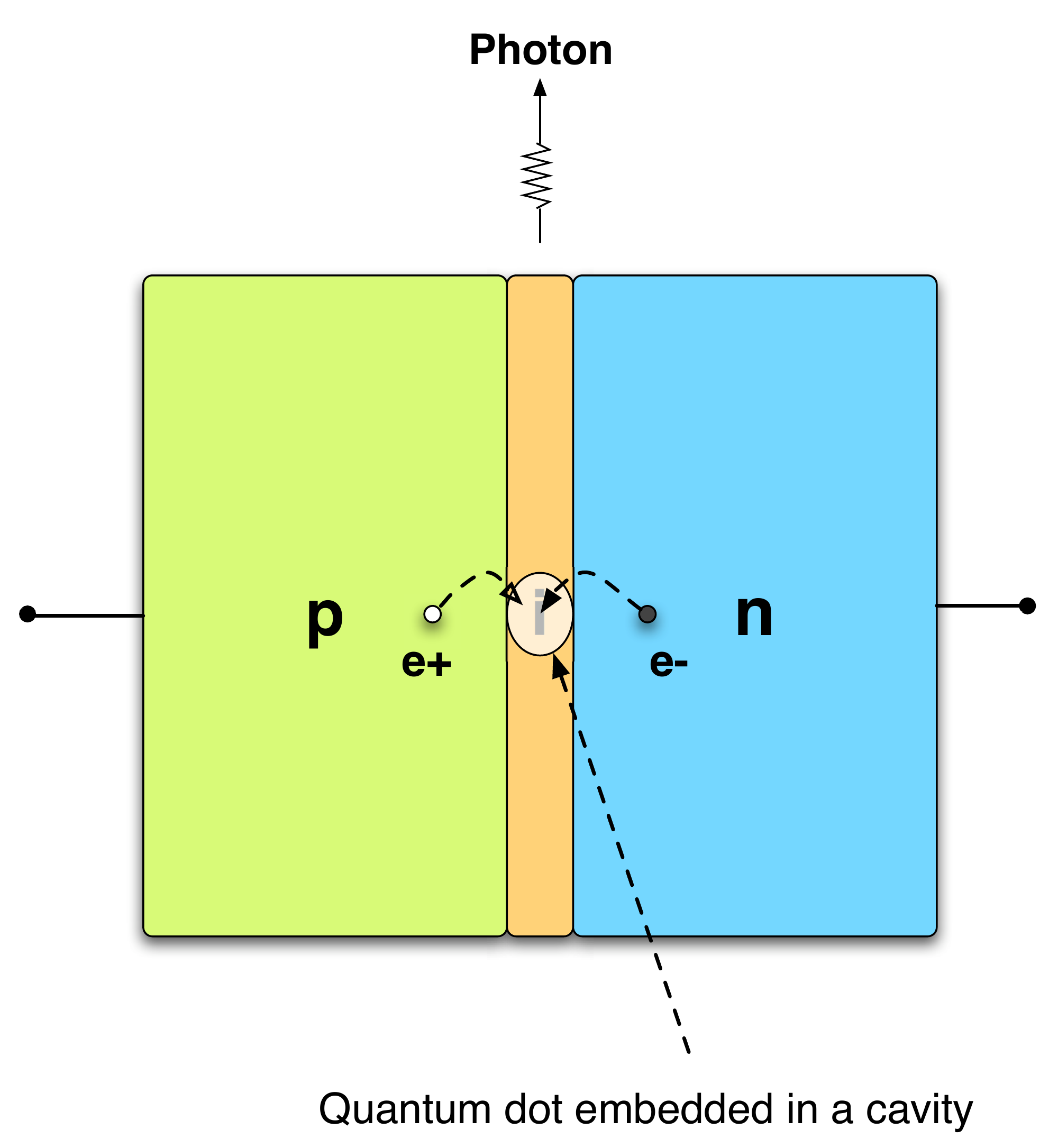}
  \end{center}
  \caption{(Color online) \textsc{Quantum dot {\it p-n} diode or {\it p-i-n} heterojunction.} A quantum dot {\it p-n} diode comprises an insulator sandwiched between {\it p}- and {\it n}-type silicon. A quantum dot is fabricated inside the insulator, and this is contained within an optical microcavity. The diode is biased in the forward direction, such that a single electron ($e^-$) tunnels through from the {\it n}-side to the dot. The electron remains in the dot until a hole ($e^+$) tunnels through to the dot from the {\it p}-side. The electron-hole pair in the dot recombines to emit a photon into the cavity, which subsequently leaks out to an external mode.}
  \label{fig:pin}
\end{figure}

 Figure \ref{fig:pin} shows a quantum dot {\it p-n} diode that acts as a single-photon source. A dot is present at the {\it p-n} junction and is assumed to be inside an optical microcavity in the weak coupling regime. We assume that the diode is forward biased and is in the Coulomb blockade regime. When biased at an appropriate level, an electron ($e^-$) tunnels through to the dot from the {\it n}-side; this electron remains in the dot until a hole ($e^+$) tunnels through from the {\it p}-side.  This leads to recombination of  $e^-$ and $e^+$, and a photon is emitted into the cavity.  The photon subsequently leaks out to an external mode, such as an outgoing waveguide.

While photon generation as described above is intuitively straightforward, we need to control the pumping so that the system generates at most one photon with high probability. Since the dot is inside a microcavity, there is a non-zero probability of multi-photon emission; this happens when there is more than one recombination event in a pumping cycle. Determining the length of a pumping cycle should ideally be determined by the knowledge of when the first e$^-$ tunneling event occurs. This is difficult to know, as the tunneling process is stochastic.  In the absence of specific information about tunneling times, the best that can be done is to time the pumping a priori, either to maximize the probability of a single photon, or keep the multi-photon probability below a given threshold.  (These are not necessarily the same thing.)  In principle, though, we can do better.

We continuously monitor the state of the dot by measuring the voltage across the heterojunction; the output record gives information about whether an $e^-$ has tunneled onto the dot.  We use this information to stop the pumping cycle when the dot is in its excited state (equivalently, after an $e^-$ tunneling event). If done correctly, the dot is excited only once, and the diode emits at most one photon. However, the measurement signal is noisy, and the decision of whether the dot is in its excited state or not is not necessarily easy. We use a well known sequential statistical decision technique known as CUmulative SUM (CUSUM), a decision circuit with a simple implementation that accommodates noisy observations~(Sec.~\ref{sec:change-detection}).  
 
\section{System modeling and parameter regime}
\label{sec:system}
%
\begin{figure*}[htp]
  \begin{center}
    	 \includegraphics[height=3.5in]{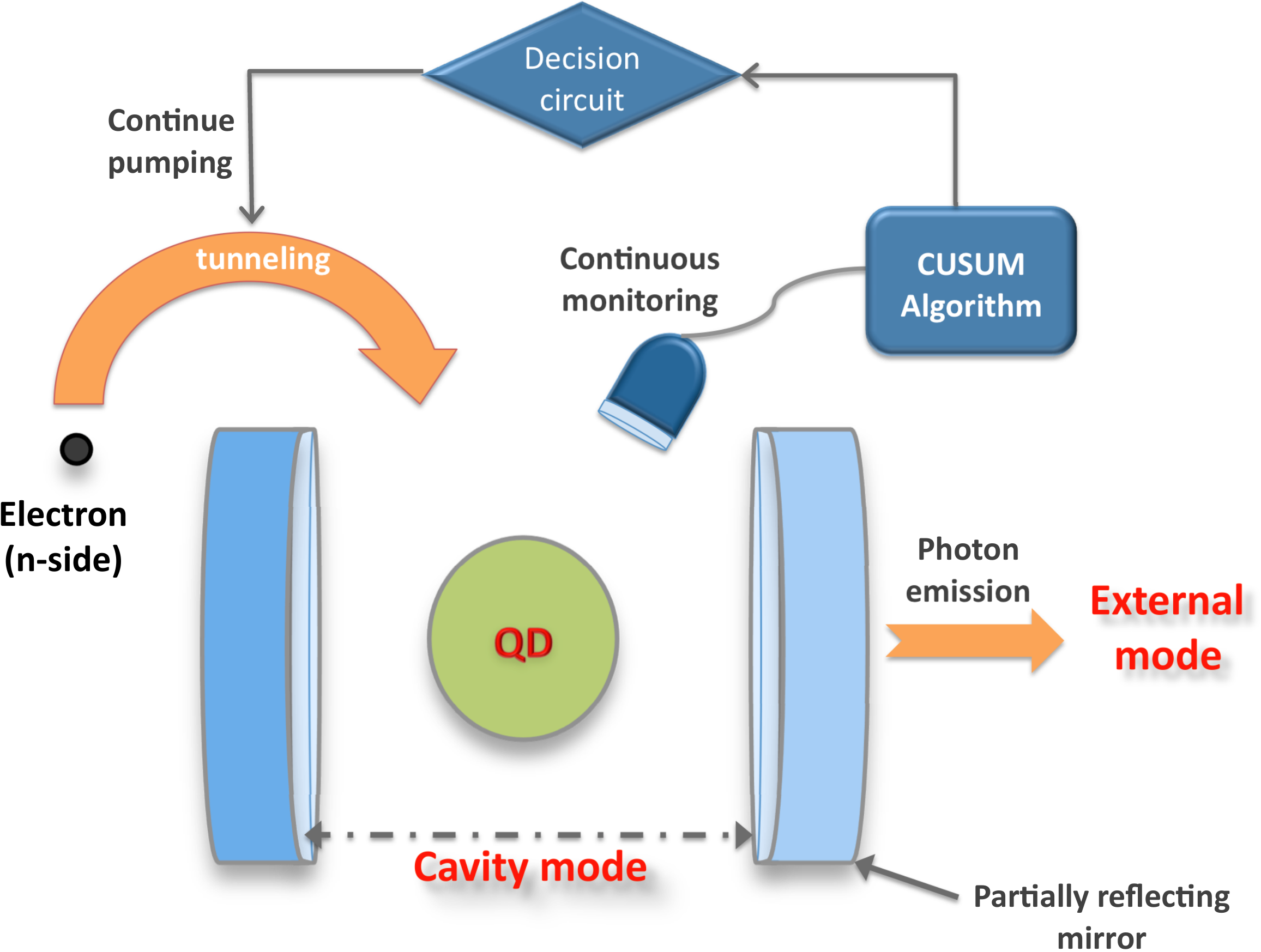}
  \end{center}
  \caption{(Color online) \textsc{Single-photon LED operation: Schematic model.}  A single-photon LED consists of three quantum objects: an emitter (QD), a microcavity and an external mode. Electrical pumping involves tunneling of an electron ($e^-$) from the {\it n}-side to the dot; an $e^- - e^+$ recombination leads to an emission of a photon into the cavity, which then leaks out to the external mode. The $e^-$ tunneling occurs when the bias voltage in the diode is favorable to the event. We control the bias voltage externally; we keep the bias `on' to allow $e^-$ tunneling, and turn it `off' to stop tunneling. We monitor the state of the dot continuously, and use a sequential algorithm called CUSUM and its associated decision circuit to decide whether to keep the bias `on' or to turn it `off.'  }
  \label{fig:schematic}
\end{figure*}
Figure~\ref{fig:schematic} schematically models the single-photon LED operation described in Sec.~\ref{sec:single-led}. The system comprises three quantum objects: a quantum dot, an optical microcavity, and an external mode (e.g., a waveguide). We assume that the dot is electrically pumped:  this allows an electron to tunnel through from the {\it n}-side to the dot, and recombine with a hole from the {\it p}-side, when the diode bias voltage is favorable. The bias voltage is controlled externally, giving us some control over the $e^-$ tunneling event.  To exert this control effectively, we continuously monitor the state of the dot and use the output to decide when to stop pumping.

We include the external mode in our description in order to calculate the various photon emission probabilities---$p(0)$, $p(1)$, etc.  These probabilities are the quantitative measure of ``goodness" of the system as a single-photon source. Obviously, we want the single-photon emission probability, $p(1)$, as close to unity as possible; unfortunately, in many parameter regimes $p(1)$ is not very close to 1, especially if it is electrically pumped. In this case we must explore a more complex trade-off landscape, as it is no longer sufficient to just maximize $p(1)$; we must also consider how high a zero-photon probability $p(0)$ and multi-photon probability $p(2+)$ we can tolerate.  For certain applications it is tolerable for the single-photon source sometimes to emit no photon, but multi-photon emission must be strongly suppressed.  Thus, in our trade-off analysis we impose an additional constraint on $p(2+)$, and try to maximize $p(1)$ subject to this constraint (see Sec.~\ref{sec:results}).

\subsection{The system}
The dot in our model has $2$ energy levels: the ground state $\ket{G}$, and the excited state $\ket{X}$. The cavity and external mode contain some number of photons, and are represented by the usual  photon number state notation---$\ket{0}$, $\ket{1}$, $\ket{2}$ etc. Thus, any state in this system has the form $\ket{G/X, 0/1/2..., 0/1/2...}$ where the order corresponds to the dot, the cavity and the external mode, respectively. 

We assume that the dot's excited state $\ket{X}$ is resonantly coupled to the cavity mode. The interaction between the dot and the cavity is given by the Jaynes-Cummings Hamiltonian: $ \H_I = \, i \hbar \, g \left( \adag \Sm \, - \, \a \Sp \right)$, where $g$ is the interaction strength,  $\adag$ ($\a$) is the creation (annihilation) operator acting on the cavity mode, and $\Sm = \ket{G}\bra{X}$; we operate in the interaction picture, and henceforth set the total Hamiltonian $\H$ to $\H_I$.  The system is initially decoupled, in the state $\ket{G,0,0}$.

Incoherent processes included in this model are pumping, spontaneous decay of the quantum dot, cavity leakage and dephasing. $\Omega$ is the rate of pumping; $\Gamma$ is the spontaneous emission rate for the $X \rightarrow G$ transition; $\kappa$ is the photon leakage rate from the cavity to the external mode, and $\gamma$ is the dephasing rate between emitter and the cavity mode. 

We treat electrical pumping as a incoherent process~\cite{electrical-incoherent}. The process of pumping involves an electron tunneling through to the dot when the bias voltage at the diode junction is favorable. Since electron hopping ($n$-side $\rightarrow$ dot) happens at random times, we model pumping as an incoherent $G \rightarrow X$ transition with rate $\Omega$.  

\begin{figure*}[htp]
  \begin{center}
    	 \includegraphics[height=3.5in]{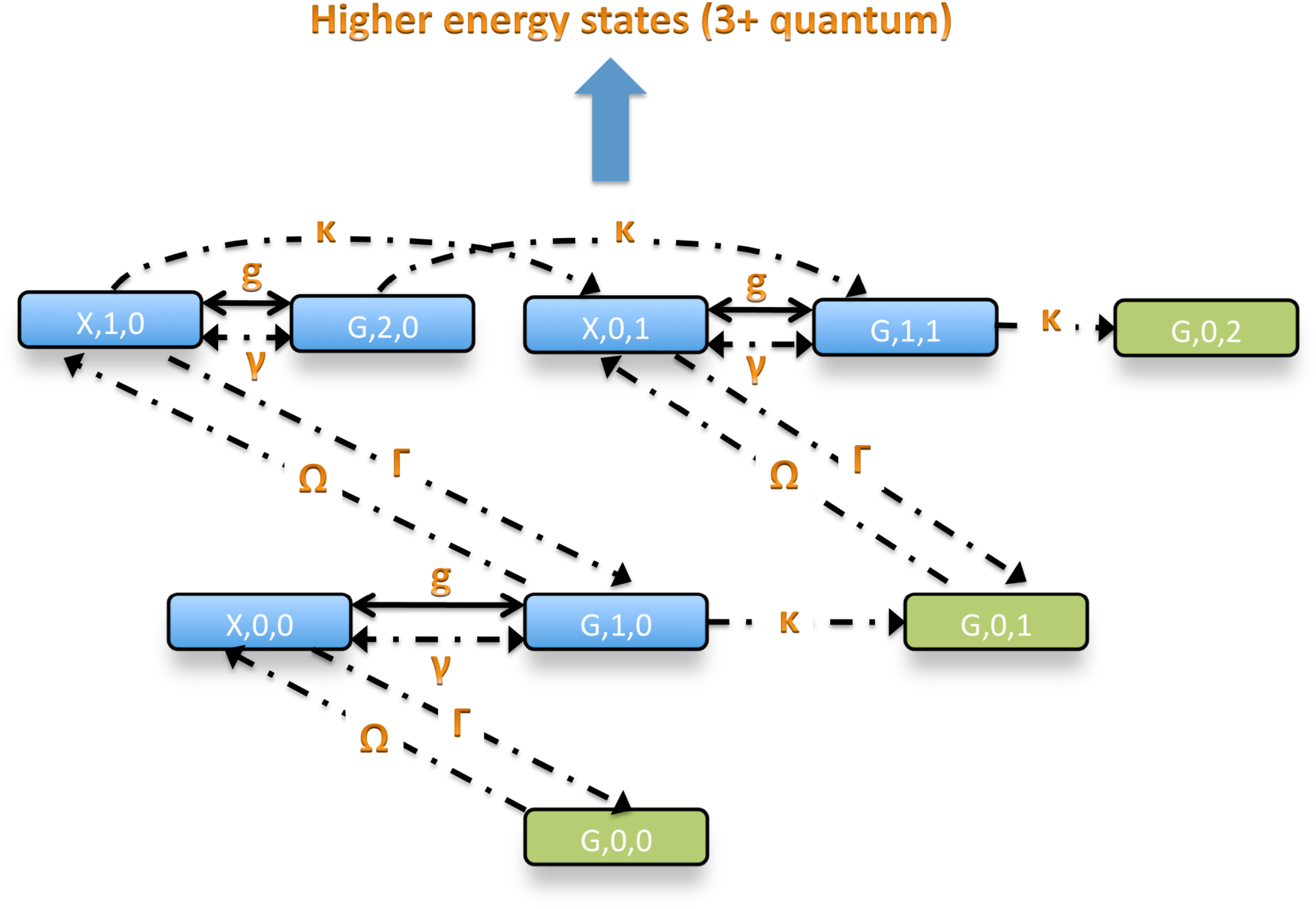}
  \end{center}
  \caption{(Color online) \textsc{Energy level diagram.}  A state of the system is described by three quantum numbers: the dot, the cavity and the external mode. In this diagram, we show those states with at most $2$ quanta of energy and their significant dynamical processes: broken arrows indicate incoherent processes, and solid arrows indicate coherent evolution. If the pumping duration is finite, $\ket{G,0,0}$, $\ket{G,0,1}$ and $\ket{G,0,2}$ are the possible final states of this system, and their corresponding probabilities represent the zero photon, $1$ photon and $2$ photon probabilities, respectively.}
  \label{fig:eld}
\end{figure*}
Figure~\ref{fig:eld} shows the energy level diagram of the single-photon LED system. The system moves up the energy ladder as we pump longer. For finite pumping time, the system eventually evolves to its possible final states $\ket{G,0,0/1/2 \dots}$, with corresponding probabilities  $p(0/1/2 \dots)$. Both the cavity mode and the external photon mode in principle have infinitely-many energy levels; however, since our goal is to generate a single-photon state $\ket{G,0,1}$ with high probability ($p(1)$ $\rightarrow 1$), and we are assuming that the coupling between the cavity and external mode is stronger than the coupling between the dot and the cavity, we truncate the higher energy states ($>2$) as shown in Fig.~\ref{fig:eld}. By truncating the state space, we make the state $\ket{G,0,2}$ now represent $\ket{G,0,2+}$---that is, a state with $2$ or more photons in it---and its corresponding probability becomes the multi-photon probability $p(2+)$. Note that this approximation neglects some effects that could in principle contribute to the single-photon probability, such as a series of reabsorptions and spontaneous emissions. Since the probability of the system being in the higher energy states is small in the first place, and the spontaneous emission rate is assumed to be quite low, the neglected effects should have little impact on the estimates of single-photon probability.

The evolution of the system is described by a stochastic master equation (SME) \cite{Trajectories}:
\begin{widetext}
\begin{equation}
d\rho \, = - \,  \frac{i}{\hbar} \, [ \H, \rho ] \, dt \, + \, \left( \Gamma \, \mathcal{H}[ \Sm]  \, +
		  \kappa \,  \mathcal{H}[ \a \bdag] \, + \, \gamma \, \mathcal{H}[ \projX] \, + \, \Omega \, \mathcal{H}[ \Sp] \, \right) \rho \, dt + \, \sqrt{\eta \gamma} \, \mathcal{D}[\projX] \,\rho \, dW_t .
\label{eqn:sme}
\end{equation}
\end{widetext}
$\mathcal{H}$ and $\mathcal{D}$ are superoperators:
\[
\mathcal{H}[\A] \rho = \A \rho \Adag - (\Adag \A \rho + \rho \Adag \A)/2,
\]
\[
\mathcal{D}[\projX]\rho = \projX \rho \, + \,  \rho \projX \, - \, 2 \expect{\projX} \rho,
\]
where $\expect{\projX}=Tr\{\projX \rho\}$ is the quantum expectation; $dW_t$ is a Brownian motion with
\[
\mathbb{E}[dW_t] \, = \, 0 \,  \text{ and } \, \mathbb{E}[dW_t dW_s] \, = \, \delta(t-s)\, ds\, dt ,
\]
which characterizes a Wiener process; here $\mathbb{E}$ is the expectation of a random variable. Also, $\Sp = \ket{X}\bra{G}$, $\bdag$ is the creation operator acting on the external mode, and $\projX = \ket{X}\bra{X}$ is the dephasing or ``observer'' operator acting on the quantum dot.  Dephasing has two potential sources:  the interaction of the emitter with other degrees of freedom, for instance phonon modes in the dot \cite{Michler-Vahala03}, and the back action of a measuring device coupled to the dot.  A measuring device will allow us to recover some information about the system of interest, but it is unlikely that we can tap into internal modes of the dot; we use $0\le\eta\le1$ to denote the measurement efficiency with which we (the ``observer'') recover information lost in dephasing.  

\subsection{Parameter regime}
We operate in the weak coupling regime, and assume that $\kappa$ is the dominant system parameter. We require that spontaneous emission $\Gamma$ be small, for higher $\Gamma$ implies lower single-photon probability. 

The technique developed in this work is based on continuous monitoring of the dot, and will be useful only when the pumping rate $\Omega$ is comparable to $g$. This is because it takes time to gather information using continuous monitoring, and with higher $\Omega$ the chances are that the decision to turn off the pumping will be too late, increasing the multi-photon probability above tolerable levels. In fact, the scenario where $g << \Omega$ has a much simpler solution: turn on pumping for time much less than the emission time scale ($ \approx  (g^{2}/\kappa)^{-1}$); this will work quite well because we move up the energy ladder (Fig.~\ref{fig:eld}) only when the dot makes multiple $X \rightarrow G$ transitions; since $\Gamma$ is very small, the transition time scale is thus set by $g$. 

We therefore are interested in the case where the parameters satisfy the following conditions:
\begin{equation}
\Gamma << \Omega,  g \le \gamma < \kappa,
\label{eqn:param_regime}
\end{equation}
for our technique to be useful.

\section{Change detection algorithm}
\label{sec:change-detection}
The output signal obtained from our continuous measurement is given (in rescaled units) by 
\begin{equation}
J(t) = \, \langle \projX \rangle(t) \,  + \, \beta \, \xi(t) 
\label{eqn:meas-rec}
\end{equation}
where  $\beta = (\eta \gamma)^{-1/2}$ and $\xi(t) = dW_t/dt$ is Gaussian white noise with zero mean, $i.e.$ $\mathbb{E}[dW_t] = 0$ and $dW_t^2 = dt$, where $\mathbb{E}$ is the expectation of a random variable  \cite{Trajectories}; $\projX = \ket{X}\bra{X}$ is the dot's excited state projection operator, while $\expect{\projX}(t)$ is its quantum expectation $Tr\{ \projX \rho(t) \}$. 

The excited state of the dot is $\ket{X}$, and the unexcited state is $\ket{G}$. The measurement output conveys the information about this state:
\begin{equation} 
 \mathbb{E}[J(t)]  =  \expect{\projX}(t) =
\left\{ \begin{array}{ll}
         0 & \mbox{if the dot is in $\ket{G}$},\\
        1 & \mbox{if the dot is in $\ket{X}$},
\end{array} \right. 
\label{eqn:mean-meas-rec}
\end{equation} 
as $\mathbb{E}[\xi(t)] = 0$. If the output contained no noise ($\beta\rightarrow0$) then the state of the dot would tend to be localized at either $\ket{X}$ or $\ket{G}$, and transitions between would be readily detectable.  The presence of noise complicates matters, since it can mask the state of the dot---indeed, if the noise is high, it is easy to miss the transitions.  Note that if we average the output signal over a short interval $\Delta t$, its variance is a constant: $\sigma^2 = \beta^2/\Delta t$.  When an electron tunnels through, there is a change in the mean of the output signal. By detecting this change, we could detect when a tunneling event takes place, and turn off the pumping. However, we cannot access the mean $\mathbb{E}[J(t)]$ directly, but have to infer it from $J(t)$ [Eq.~(\ref{eqn:meas-rec})].  

We obtain the output signal $J(t)$ continuously in time, and the decision to turn off the pumping must be made in real time. This imposes practical restrictions on the kind of algorithms that are feasible; for instance, Bayesian machine learning-type algorithms~\cite{stat-decision} may be too slow to be useful in practice.  Sequential algorithms~\cite{stat-decision} are procedures that use only the output signals gathered to the present time, and not a priori information, to infer the probability density function (pdf).  As such, they are generally suboptimal, and will succeed only in certain parameter regimes, but they have the virtue of being easier to implement in practice.  One such algorithm is Cumulative Sum (CUSUM). 

\subsection{Sequential CUSUM procedure}
\label{sec:cusum}

We present a discrete version of CUSUM here for simplicity, which can straightforwardly be extended to the continuous case. Let $y_1 = J(t_1), y_2 = J(t_2), \dots$ represent the time series of the output signal, averaged over a succession of intervals of size $\Delta t$. We know that $Y_k$ is a $2$ parameter Gaussian random variable with a variable mean ($\mu$) and a constant variance ($\sigma^2 = \beta^2/\Delta t$). We know from Eq.~(\ref{eqn:mean-meas-rec}) that if the dot is localized onto a single energy state, the mean has two possible values: $\mu_0 = 0$ or $\mu_1 = 1$. Detecting a change in the mean is therefore equivalent to a simple hypotheses testing problem:
\begin{equation}
\left. \begin{array}{c}
H_0: \mu = \mu_0 \\
H_1: \mu = \mu_1
\end{array} \right\}.
\label{eqn:hypotheses}
\end{equation}

CUSUM is based on a sequential probability ratio test (SPRT). SPRT, in turn, is based on the concept of log-likelihood ratios.  These are defined by
\begin{equation}
S_n  = S(y_1^n) = \ln\left( \frac{p_{\mu_1}(y_1^n)}{p_{\mu_0}(y_1^n)} \right),
\end{equation}
where $p_{\mu_0,\mu_1}(y)$ is the probability density function (pdf) given means $\mu_0$ and $\mu_1$, respectively; $S$ is called the {\it sufficient statistic} in the parlance of statistics~\cite{stat-decision}; note $y_1^n$ includes all output signals ($y_1 \dots y_n$). Further, assuming that the random variables $Y_1, \dots ,Y_n$ are independent and identically distributed (iid), 
\begin{equation}
S_n = \sum_{k =1}^n \, \ln \left(  \frac{p_{\mu_1}(y_k)}{p_{\mu_0}(y_k)} \right)  =  \sum_{k=1}^n s_k \,;
\label{eqn:Sn}
\end{equation}
$s_k$ is the sufficient statistic for random variable $Y_k$. Utilizing the fact that $y_k$ is Gaussian with pdf
\[
p_{\mu_{0/1}}(y_k) = \frac{1}{\sqrt{2 \pi \sigma^2}} \, \exp\left( - \frac{(y_k - \mu_{0/1})^2}{2 \sigma^2} \right),
\]
we have
\begin{eqnarray}
s_k &=& \frac{\mu_1 - \mu_0}{\sigma^2} \left( y_k - \frac{\mu_0 + \mu_1}{2} \right) \nonumber\\
&=& \frac{\mu_1 - \mu_0}{\beta^2} \left( y_k - \frac{\mu_0 + \mu_1}{2} \right)\Delta t  .
\label{eqn:sk}
\end{eqnarray}

Since the bias voltage is on at the start of a pumping cycle, and the algorithm has to decide when to turn it off, CUSUM has to detect the change $\mu_0 \rightarrow \mu_1$. Ignoring the constant pre-factor in Eq.~(\ref{eqn:sk}), it is easy to see that $S_n$ [Eq.~(\ref{eqn:Sn})] has a negative drift if the dot is in its ground state ($y_k \approx \mu_0 = 0$), and a positive one when in excited state ($y_k \approx \mu_1 = 1$); a tunneling therefore should cause a V-shaped profile in $S_n$. As the signal is noisy, we use an appropriate threshold to mitigate false positives. 

To this end, we define 
\begin{equation}
m_k = \min_{1 \le j \le k} \; S_j.
\end{equation}
We calculate $m_k$ using a simple procedure: at each step, set $m_k = S_k$ if $S_k < m_{k-1}$, otherwise it retains its previous value.  In hardware, this can be done with a single register and a comparator. The decision rule, at each time step, is
\begin{equation}
d_k = 
\left\{ \begin{array}{ll}
	H_0 & \mbox{if $(S_k \, - \, m_k) \le h$}, \\
	H_1 & \mbox{otherwise},
\end{array} \right. 
\label{eqn:decision-rule}
\end{equation} 
where $h$ is the threshold, chosen based on the parameters of the system to mitigate false positives. We stop pumping if $d_k = H_1$.

\subsection{Bayesian solution}
\label{sec:bayesian}

\begin{figure}[htp]
  \begin{center}
    	 \includegraphics[height=3.0in]{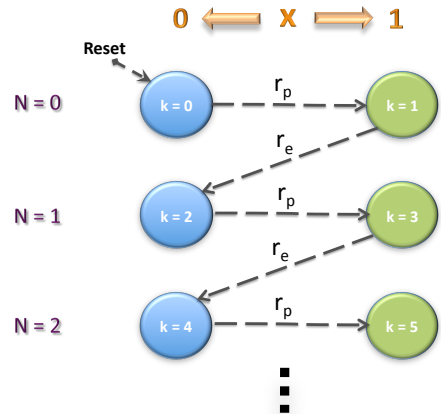}
  \end{center}
  \caption{(Color online) \textsc{Markov chain approximation.} We approximate the energy level diagram as a Markov chain. The individual states represent the state of the dot along with some number of photons in the external mode. The state of the dot is represented by $x$, and $n$ is the number of photons in the external mode; $x$ is $0$ if there is no $e^-$ in the dot and $1$ if an $e^-$ is present; $n$ is a positive integer. The parameters $r_p$ and $r_e$ are the pumping and emission rates, respectively. We combine $x$ and $n$ into a single variable $k = 2 n \, + \, x$; note that odd $k$ states contain an $e^-$ while even $k$ states have no $e^-$ in the dot. The system starts initially in the state $k=0$ with no $e^-$ in the dot and no photon in the external mode.   }
  \label{fig:mc}
\end{figure}

The primary difference between a Bayesian and sequential solution is the use of prior probabilities in the former case. We approximate the system described in Fig.~\ref{fig:eld} as a Markov chain, shown in Fig.~\ref{fig:mc}. This approximation is good because the system is assumed to operate in the weak coupling regime [Eq.~(\ref{eqn:param_regime})], where the cavity decay rate $\kappa$ dominates the emitter$-$cavity coupling strength $g$.

A state in the Markov chain represents the state of the dot and the external mode. The state of the dot is represented by $x$, and takes values $0$ or $1$ depending on whether an electron has tunneled through to the dot or not, with $1$ indicating the presence of an electron. The external mode contains $n$ photons, where $n$ is a non-negative integer. We combine the state of the dot ($x$) and cavity ($n$) into a single variable using,
\begin{equation}
k \: = \: 2 \, n \, + \, x.
\label{eqn:k}
\end{equation}
Even $k$ indicates the absence of an electron, and odd $k$ the presence of an electron in the dot. The parameters $r_p$ and $r_e$ are the pumping and effective emission rates, respectively; here, $r_p = \Omega$ and $r_e = 4 g^2/\kappa$. (We ignore spontaneous emission $\Gamma$ in this approximation as its contribution is small, see Eq.~(\ref{eqn:param_regime}).)

Let $p_k(t)$ be the probability of state $k$ at time $t$. The system starts in state $k=0$, so $p_0(0) = 1$ and $p_k(0)=0$ for all $k > 0$. The Markov chain in Fig.~\ref{fig:mc} can be described by a set of coupled differential equations:
\begin{eqnarray}
\frac{d p_0}{dt} \, &=& \, - \, r_p \, p_0  \nonumber \\
\frac{d p_1}{dt} \, &=& \, r_p \, p_0 \, - \, r_e \, p_1 \\
\frac{d p_2}{dt} \, &=& \, r_e \, p_1 \, - \, r_p \, p_2 \nonumber \\
\vdots \nonumber
\label{eqn:prior-prob}
\end{eqnarray}
Using these initial conditions, we can solve the system of equations analytically; we first solve for $p_0$, then use this solution to solve for $p_1$, and so on.

Equation~(\ref{eqn:prior-prob}) describes the Markov chain in Fig.~\ref{fig:mc} when no measurement output is available. To incorporate the information from the continuous measurements, we first consider a fixed time step $\Delta t$, and then go to the continuum limit. During each $\Delta t$ the system evolves according to Eq.~(\ref{eqn:prior-prob}); then at the end of the interval $\Delta t$ we update the probabilities by conditioning on the measured output. We now derive update formulas (conditional probabilities) for $p_k$'s using the Bayes rule.

Let
\begin{equation}
y(t + \Delta t) = \expect{\projX}(t + \Delta t ) \,  \Delta t \, + \, \beta \, \Delta W_{t + \Delta t }, \label{eqn:y-t}
\end{equation}
where $W_{t + \Delta t}$ is a Wiener process with $\mathbb{E}[\Delta W_{t+\Delta t}] = 0$ and $\mathbb{E}[\Delta W_{t+\Delta t}^2] = \Delta t$; $\expect{\projX}(t + \Delta t )$ is the quantum expectation of the state of the dot. We define
\begin{eqnarray}
Q_0 &\equiv& p(x = 0) \, = \, p_0 \, + \, p_2 \, +  \hdots ,    \\
Q_1 &\equiv& p(x = 1) \, = \, p_1 \, + \, p_3 \, + \hdots  ,
\label{eqn:q0-q1}
\end{eqnarray} 
where $Q_1$ and $Q_0$ represent the probabilities of an electron  to be present in the dot or not (corresponding to $x$ being $1$ or $0$). They obey the simple equations
\begin{eqnarray}
d Q_0 &=& (- \, r_p \, Q_0 \, + \, r_e \, Q_1) \, dt , \\
d Q_1 &=& (+\,r_p \, Q_0 \, - \, r_e \, Q_1) \, dt ,
\label{eqn:dq0-dq1}
\end{eqnarray}
with initial conditions $Q_0(0) = 1$ and $Q_1(0) = 0$.  We have used Eq.~(\ref{eqn:prior-prob}) in the above derivation.

Since $y$ is Gaussian random variable with means $0/\Delta t$, we have
\begin{eqnarray}
p(y|x=0) \, &=& \, \frac{1}{\sqrt{2 \pi \sigma^2}} \, \exp{\{- y^2/2 \sigma^2\}}, \\
p(y|x=1) \, &=& \, \frac{1}{\sqrt{2 \pi \sigma^2}} \, \exp{\{- (y - \Delta t)^2/2 \sigma^2\}},
\end{eqnarray}
where, $\sigma^2 = \mathbb{E}[y^2] \approx \beta^2 \Delta t$ is its variance.  Also,
\begin{eqnarray}
p(y) \, &=& \, p(y|x=0) \, p(x=0) \, + \, p(y|x=1) \, p(x=1) \nonumber \\
&=& \frac{1}{\sqrt{2 \pi \sigma^2}} \left(  Q_0 \exp{\{-y^2/2 \sigma^2\}} \, + \right. \nonumber \\
&& \; \;\;\;\; \;\;\;\; \;\;\;\; \;\;\;  \left. Q_1 \exp{\{ - (y-\Delta t)^2/2\sigma^2\}} \right) ,
\end{eqnarray}
so that
\begin{eqnarray}
\frac{p(y|x=0)}{p(y)}  &=&  \left[ Q_0  +  Q_1  \exp{\left\{\frac{y^2 - (y - \Delta t)^2}{2 \sigma^2} \right\}  } \right]^{-1}, \nonumber 
\end{eqnarray}
which using the relation $\sigma^2 = \beta^2 \Delta t$ becomes
\begin{eqnarray}
&=& \left[ Q_0 + Q_1 \exp{ \left\{ \frac{y}{\beta^2} \right\} } \exp{ \left\{ \frac{- \Delta t}{\beta^2} \right\} } \right]^{-1}. \nonumber
\end{eqnarray}
Assuming that $\Delta t \ll \beta^2$, we get 
\begin{eqnarray}
&\approx& \left[ Q_0 + Q_1\left( 1 + \frac{y}{\beta^2} + \frac{\Delta t}{2\beta^2}\right)\left(1 - \frac{\Delta t}{2 \beta^2} \right) + O(\Delta t^{3/2}) \right]^{-1} \nonumber \\
&\approx& \left[ 1 \, + \, Q_1 \left( \frac{y}{\beta^2} \right) + O(\Delta t^{3/2}) \right]^{-1} \nonumber \\
&\approx& 1 \, - \, Q_1 \left( y \, - \, Q_1 \Delta t \right)/\beta^2. 
\label{eqn:update-x0}
\end{eqnarray}
In the above derivation, we used the fact that the total probability is conserved, and hence $Q_0 + Q_1 = 1$ at all times. Similarly,
\begin{eqnarray}
\frac{p(y|x=1)}{p(y)}  &\approx& 1 \, + \, Q_0 \left( y \, - \,(1 - Q_0) \Delta t \right)/\beta^2.
\label{eqn:update-x1}
\end{eqnarray}
Therefore the update formula for $Q_0$ is
\begin{eqnarray}
Q_0 &\rightarrow& \, \left( \frac{p(y|x=0)}{p(y)} \right) Q_0 \nonumber \\
&=& Q_0 \, - \, Q_0 Q_1 ( y \, - \, Q_1 \Delta t)/\beta^2
\label{eqn:update-q0}
\end{eqnarray}
and that of $Q_1$ is
\begin{eqnarray}
Q_1 &\rightarrow&\,  \left( \frac{p(y|x=1)}{p(y)} \right) Q_1 \nonumber \\
&=& Q_1 \, + \, Q_0 Q_1 ( y \, - \, (1 - Q_0) \Delta t)/\beta^2. 
\label{eqn:update-q1}
\end{eqnarray}

We now combine the system dynamics given by Eq.~(\ref{eqn:dq0-dq1}) with the update formulae due to observation from Eqs.~(\ref{eqn:update-q0}) and (\ref{eqn:update-q1}) to get
\begin{eqnarray}
Q_1(t &+& \Delta t) \, = \, Q_1(t) \, - \, (r_p + r_e) \,  Q_1(t) \, \Delta t \, + \, r_p \, \Delta t \nonumber \\
&& + \, Q_1(t) (1 \, - \, Q_1(t)) \left( \frac{y(t) \, - \, Q_1(t) \Delta t}{\beta^2} \right).
\end{eqnarray}
We can infer $Q_0$ straightforwardly using $Q_0 + Q_1 = 1$. Since $x$ only takes values $0$ and $1$, its expected value $\overline{x}$ is
\begin{equation}
\overline{x} = \mathbb{E}[x] = 1\times p(x=1) = Q_1.
\label{eqn:xbar}
\end{equation}

We can now make precise the question of when to stop pumping. We need to calculate the individual probabilities $p_k$ including the prior probabilities [Eq.~(\ref{eqn:prior-prob})] and the update formula due to continuous monitoring of the dot. Here, we use $(n,x)$ notation instead of $k$ to represent a state in the Markov chain (Fig.~\ref{fig:mc}), for clarity. Applying the Bayes rule we get
\begin{eqnarray}
p(n,x|y) \, &=& \, \frac{p(y|n,x)}{p(y)} \, p(n,x) \, = \, \frac{p(y|x)}{p(y)} \, p(n,x).
\label{eqn:p-nx}
\end{eqnarray}
We have used the fact that the observation $y$ is independent of number of photons $n$ in the external mode. Switching back to the variable $k$ we get
\begin{equation}
p(k|y) \, = \, \frac{p(y|x)}{p(y)} \, p(k).
\label{eqn:k-update}
\end{equation}
Using the above relation with update formulas in Eqs.~(\ref{eqn:update-x0}) and (\ref{eqn:update-x1}), along with prior probabilities given in Eq.~(\ref{eqn:prior-prob}), we get 
\begin{eqnarray}
p_0(t &+& \Delta t ) \, = \, p_0(t) - \, r_p \, p_0(t) \, \Delta t \nonumber \\
&& \,\, - \, p_0(t) \, \left( \frac{\overline{x}(t) \, (y \, - \, \overline{x} \, \Delta t)}{\beta^2} \right),  \label{eqn:p0-t} \\
p_1(t &+& \Delta t ) \, = \,p_1(t) - \,(r_e \, p_1(t) \, - \, r_p \, p_0(t)) \, \Delta t \nonumber \\
&& + \, p_1(t) \, \left( \frac{(1 \, - \, \overline{x}(t)) \, (y(t) \, - \, \overline{x}(t) \, \Delta t)}{\beta^2} \right),  \label{eqn:p1-t} \\
p_2(t &+& \Delta t) \, = \,  p_2(t) \, - \, ( r_p \, p_2(t) \, - \, r_e \, p_1(t)) \,  \Delta t \nonumber \\
&& \, - \, p_2(t) \, \left( \frac{\overline{x}(t) \, (y(t) \, - \, \overline{x}(t) \, \Delta t)}{\beta^2} \right),  \label{eqn:p2-t} \\
\vdots \nonumber
\end{eqnarray}
Here, $\beta$ represents the quality of measurement, effectively the inverse of the signal-to-noise ratio (SNR):  the higher the value of $\beta$, the lower the signal quality. Note that as $\beta$ is reduced (better signal quality), the terms from the (Bayesian) update formula dominates, indicate that our estimate of the probabilities $p_k$ derives mostly from our observation; while as $\beta \rightarrow \infty$ the probability $p_k$ converges to its a priori solution Eq.~(\ref{eqn:prior-prob}). Therefore, the Bayesian technique should perform {\it no worse} than the a priori solution, at least as long as the Markov chain approximation remains good.

Observe that states $k \in (0,1,2)$ can emit at most $1$ photon, while states $k > 2$ emit $2$ or more photons. If we bound the tolerable multi-photon probability by $\epsilon$, then our decision circuit becomes straight-forward: we continue to pump until $p(k>2) \, = \, p_3 + \, p_4 \, + \, \hdots \,  <  \, \epsilon$, and stop pumping as soon as the inequality is violated. The above condition can be re-written in more convenient form: 
\begin{equation}
p(k \le 2) \, = \, p_0 \, + \, p_1 \, + \, p_2 \, \ge \, (1 \, - \, \epsilon),
\end{equation}
and as before, we stop pumping when the inequality is violated. In this form it suffices to keep track of just $3$ probabilities---$p_0, p_1,p_2$---together with the expectation value $\bar{x}$.

\section{ Results } 
\label{sec:results}
In Sec.~\ref{sec:change-detection}, we presented two different approaches---sequential and Bayesian---to improve single-photon probability in the presence of continuous monitoring. The Bayesian solution (Sec.~\ref{sec:bayesian}) is computationally expensive, but provides a smooth transition from low-noise limit to the high-noise one; since in the high-noise limit the Bayesian updates converges to an a priori solution, we will do no worse than the deterministic solution (where no measurement is done). Sequential CUSUM (Sec.~\ref{sec:cusum}) on the other hand, is simple and requires less computational resources (an integrator, a comparator and $2$ registers). This is important because the decision to turn off the pumping has to be taken in real time. Though the Bayesian solution is very useful for the insight it provides, particularly in high-noise settings, in this section we explore sequential CUSUM, as it is far easier to implement in real time.  At high SNR it should approach the performance of the Bayesian solution.

We first discuss the numerical values of the system parameters. Then we consider the benchmark against which we compare the CUSUM technique:  the a priori evolution of Eq.~(\ref{eqn:prior-prob}) without continuous monitoring (Fig.~\ref{fig:det}). We compare the results of CUSUM to this deterministic solution, and plot the single-photon probability as a function of pumping rate $\Omega$ (Fig.~\ref{fig:cusum}); we consider $3$ cases in CUSUM corresponding to different measurement quality ($\beta$) regimes: ($i$) low-noise, ($ii$) intermediate-noise, and ($iii$) high-noise.  The monitoring efficiency $\eta$ has a strong effect on on the performance of the CUSUM algorithm, and we explore this dependence.  For low $\eta$ ($= 0.1$), CUSUM is not useful, and in fact is detrimental.  In this regime, the measurement output is so noisy that CUSUM cannot recognize when the dot becomes excited.  In such a case, increasing the monitoring strength (and hence the dephasing rate) nominally (and thereby improving $\eta$) can lead to regimes where the technique performs better (Fig.~\ref{fig:deffs}). Note that in all our simulations, we use the constraint 
\begin{equation}
\text{p($2+$)} \, \le \, 1\%, 
\label{eqn:p2+}
\end{equation}
That is, we require that all solutions satisfy the condition that the multi-photon probability can at most be $1\%$.  We use the fourth-order Runge-Kutta integrator $rk4$ \cite{Press07} to numerically integrate the stochastic master equation~(\ref{eqn:sme}). 

%
\subsection{Parameter values}
\label{sec:params}
In our model, five parameters characterize the system: $g$, $\Omega$, $\gamma$, $\Gamma$, and $\kappa$. We rescale all the parameters with respect to the cavity decay rate $\kappa$, which establishes the dimensionless (frequency) units for the simulation. In these dimensionless units, the parameter values are:
\begin{equation}
 	g = 0.1, ~ \Gamma = 0.001 ~ \text{and} ~ \kappa = 1.0.
\label{eqn:parameters}
\end{equation}
The value of $\kappa$ in physical (frequency) units is $\sim 95$ KHz (in \cite{Heindel10}, $f = 220$ MHz and $Q = 2300$; in frequency units, $\kappa = f/Q$) . The physical units of other parameters can be obtained straightforwardly by rescaling with respect to $\kappa$. 

Pumping rate $\Omega$  and measurement strength $\gamma$ are interesting from our standpoint, in that the effectiveness of CUSUM as compared to an a priori strategy is strongly affected by them; $\Omega$ determines the time-window for the decision circuit while measurement quality ($\beta = (\gamma \eta)^{-1/2}$) influences the ability to make the right decision, that is, to turn off the pumping at the right time.  The a priori solution also improves with higher $\Omega$, further eroding the benefits of the CUSUM protocol.

%
\subsection{Deterministic solution}
\label{sec:det}
\begin{figure}[htp]
  	\includegraphics[width=3.25in,height=2.75in]{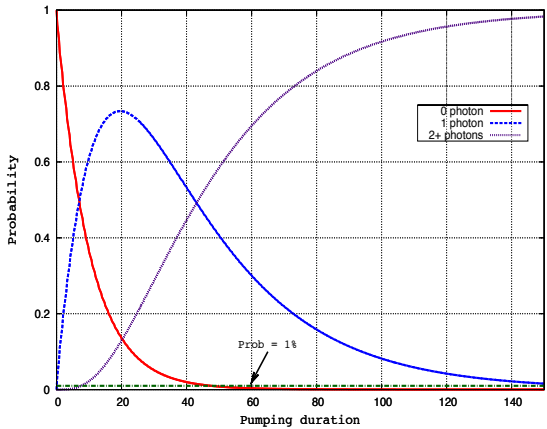} 
  \caption{(Color online) \textsc{Deterministic evolution.}  We plot photon probabilities---$p(0)$, $p(1)$ and $p(2+)$---as a function of the pumping duration; the parameter values are given in Eq.~(\ref{eqn:parameters}) and $\Omega = 0.1$. The system is initially assumed to be in the state $\ket{G,0,0}$. Initially $p(0) = 1$ and decreases as pumping duration is increased; $p(1)$, on the other hand, starts at $0$, increases initially, reaches a maximum and then starts to decrease; $p(2+)$ increases with pumping duration (more gradually than $p(1)$) and starts to dominate at long times. The maximum of $p(1)$ is $\approx 0.73$ and it does so at $t\sim 19.5$; however, $p(2+)$ is about $12\%$, which is quite high. Imposing the constraint in Eq.~(\ref{eqn:p2+}), the best time to stop pumping at these parameter values is $t\sim8$, where $p(1)\approx 53\%$ and $p(0)\approx 46\%$.  }
  \label{fig:det}
\end{figure}

To evolve the system deterministically, we integrate $\mathbb{E}[ d \rho]$ where $d \rho$ is defined in Eq.~(\ref{eqn:sme}); due to the expectation $\mathbb{E}[.]$, the stochastic contributions vanish and the equation reduces to the usual deterministic Lindblad master equation \cite{Trajectories}. 

Figure \ref{fig:det} shows the photon number probabilities---$p(0)$, $p(1)$ and $p(2+)$---as a function of pumping duration for parameters values defined in Eq.~(\ref{eqn:parameters}) and $\Omega=0.1$.  The probabilities are obtained by integrating the Lindblad master equation with pumping turned on up to time $t$, and continuing the simulation for a sufficiently long time after the pumping is turned off.  The initial state is $\ket{G,0,0}$. We see that $p(0)$ falls exponentially as the pumping time is increased, while $p(1)$ increases initially, reaches a maximum, and then falls off; $p(2+)$ increases slowly but grows to 1 at long times. The maximum of $p(1)$ is about $73\%$ at $t\sim20$; however, p($2+$) is about $12\%$, which is unacceptably high. Imposing the constraint Eq.~(\ref{eqn:p2+}), we find that the best solution is  $p(1)\approx 53\%$, and is achieved for a pumping time $t\sim8$.

\subsection{CUSUM performance}
\label{sec:perform}
\begin{figure*}[htp]
	\includegraphics[height=3.5in]{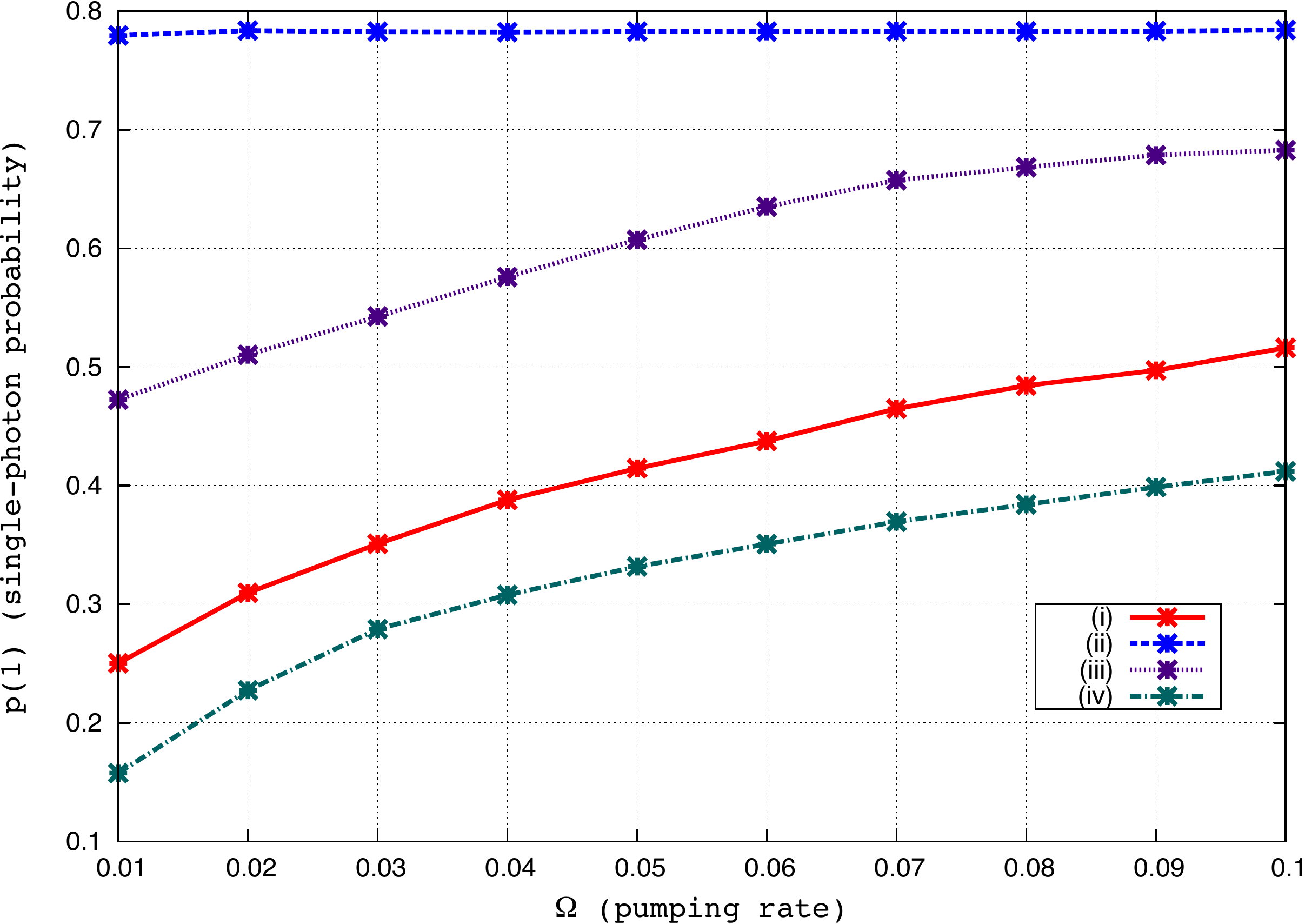}
  \caption{(Color online) \textsc{CUSUM performance.} We plot the best $p(1)$ given constraint Eq.~(\ref{eqn:p2+}) for different pumping rates $\Omega \in [ 0.01, 0.1]$; we find a close-to-optimal value for the threshold $h$ by numerical exploration. There are $4$ cases corresponding to different measurement qualities $\beta = (\gamma \eta)^{-1/2}$ (here $\eta = 1$): ($i$) deterministic (no measurement), ($ii$) low-noise ($\gamma = 10$), ($iii$) intermediate noise ($\gamma = 1.0$), and ($iv$) high-noise ($\gamma = 0.1$); we plot the deterministic case for comparison. Low-noise ($ii$) naturally leads to the best performance while high-noise ($iv$) performs the worst; in fact, ($iv$) performs worse than the deterministic case ($i$); in the intermediate regime ($iii$), CUSUM does better than cases ($i$) and ($iv$). In cases ($ii$) and ($iii$), the performance improvement is higher for lower $\Omega$ and the improvement reduces as $\Omega$ is increased. Physically, lower pumping rate means that the successive e$^-$ tunneling events are spread out in time, thus giving the CUSUM decision circuit has longer time to make the right decision; for higher rates, this time-window for CUSUM decision circuit is diminished, leading to lesser performance improvement.  Also, the deterministic solution is better for higher $\Omega$, leaving less room for improvement by CUSUM. }
  \label{fig:cusum}
\end{figure*}

We analyze the performance of CUSUM by integrating the stochastic master equation~(\ref{eqn:sme}) for the parameter values in Eq.~(\ref{eqn:parameters}), with $\Omega \in [0.01, 0.1]$; we set $\eta = 1$, and impose Eq.~(\ref{eqn:p2+}). We assume that the system is initially decoupled and starts in $\ket{G,0,0}$. The decision rule given in Eq.~(\ref{eqn:decision-rule}) is used to stop pumping.  This rule requires us to specify the threshold value $h$, to avoid false positives; we find a close-to-optimal value for $h$ by numerical exploration. For each $\Omega$, we continue to integrate each trajectory for sufficiently long time after the decision circuit stops pumping to calculate the photon probabilities.  We repeated this procedure for $1000$ trajectories and averaged to obtain $p_1(\Omega)$.  We plot this in Figure~\ref{fig:cusum}.

In Fig.~\ref{fig:cusum}, we consider $4$ cases corresponding to different measurement quality $\beta = (\gamma \eta)^{-1/2}$: ($i$) deterministic (no measurement), ($ii$) low-noise $\gamma = 10$, ($iii$) intermediate noise $\gamma = 1$, and ($iv$) high-noise $\gamma = 0.1$. The better the quality of measurement (lower $\beta$), the more accurate is our knowledge of the state of the dot; naturally, we expect CUSUM to perform well for lower $\beta$ and to fare badly when $\beta$ is high. This is borne out in Fig.~\ref{fig:cusum} where we observe case ($ii$) performing the best and ($iv$) the worst. In fact, ($iv$) does worse than the deterministic case; this means that we are better-off not using CUSUM in the high-noise limit.  

Observe that cases ($ii$) and ($iii$) provides greater improvement for lower $\Omega$, and the improvement reduces as $\Omega$ is increased. Physically, a smaller pumping rate means that the successive e$^-$ tunneling events are spread out in time, giving sufficient time for the decision circuit to make the right decision; higher $\Omega$ means less time between successive e$^-$ tunneling events, and consequently a tighter time-window for the decision circuit.  Also, as we have discussed above, for high $\Omega$ the optimal strategy is strong pumping for a short time, and we do not expect monitoring and feedback to yield much improvement.

\subsection{Measurement Efficiency}
\label{sec:efficiency}
%
The results in the previous section assumed that the monitoring efficiency was $\eta = 1$. This is unrealistic even in principle, because any realistic quantum dot will have multiple sources of dephasing, and we cannot have access to information that is lost to the dot's internal degrees of freedom.  On top of this difficulty in principle, efficient monitoring is hard to do in practice. An obvious question is: how does lower $\eta$ affect performance?

\begin{figure}[htp]
	\begin{center}
		\includegraphics[width=3.25in,height=2.75in]{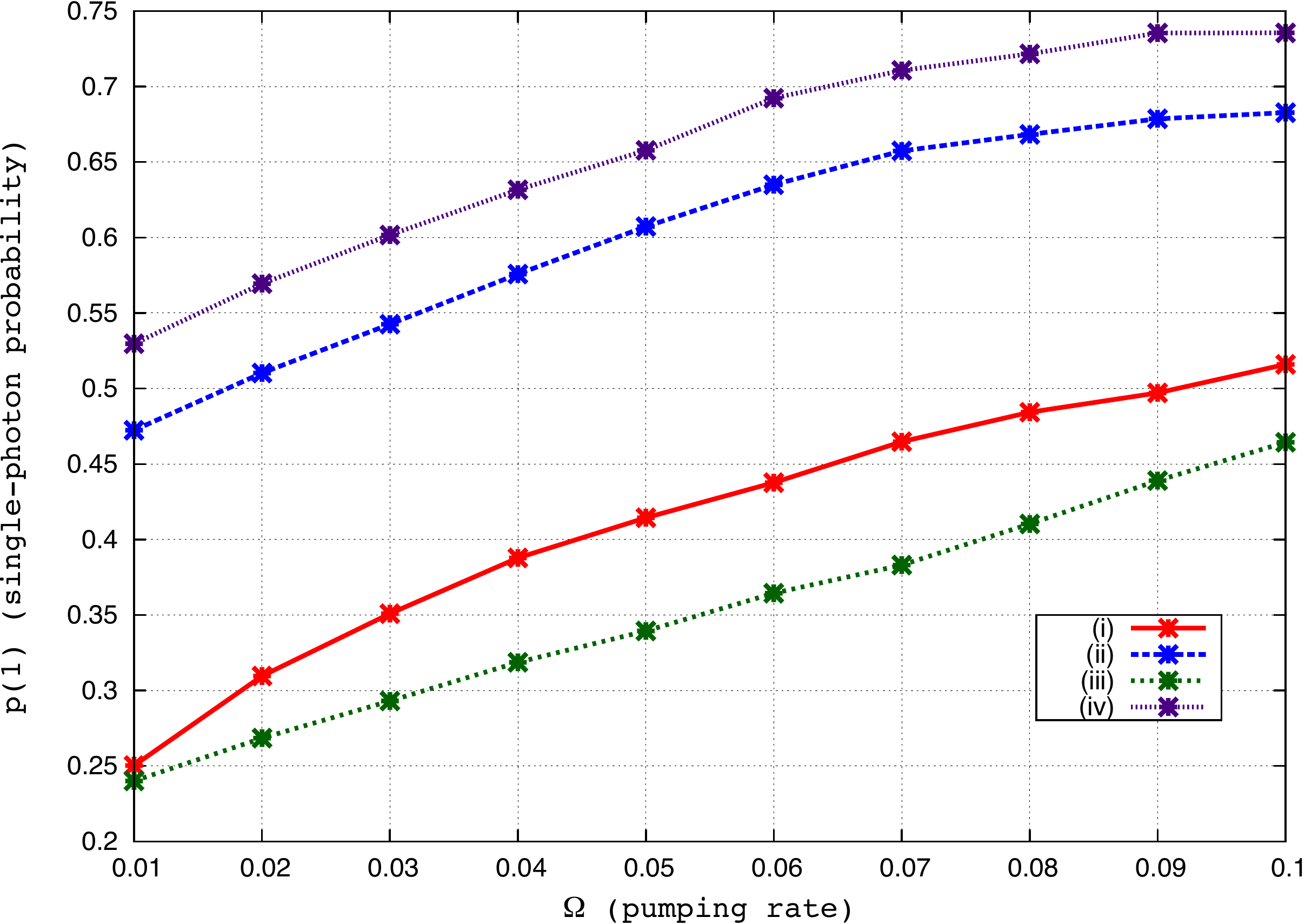}
	\end{center}
	\caption{(Color online) \textsc{Monitoring efficiency and performance.} In this plot, we explore the effect of monitoring efficiency on performance. The $4$ cases considered are: ($i$) ($\gamma=1.0, \eta = 0$), ($ii$) ($\gamma = 1.0, \eta = 1.0$), ($iii$) ($\gamma = 1.0, \eta = 0.1$), and ($iv$) ($\gamma = 2.0, \eta = 0.5$). Case ($i$) performs consistently better than ($iii$) while ($ii$) dominates both ($i$) and ($iii$). However, case ($iv$) performs even modestly better than ($ii$). Hence, a small increase in monitoring strength can compensate for lower efficiency. }
	\label{fig:deffs}
\end{figure}

To understand the effect of $\eta$ we simulated $4$ cases: ($i$) deterministic ($\gamma=1.0, \eta = 0$), ($ii$) low monitoring strength but high efficiency ($\gamma = 1.0, \eta = 1.0$), ($iii$) low monitoring strength and low efficiency ($\gamma = 1.0, \eta = 0.1$), and ($iv$) higher monitoring strength and moderate efficiency ($\gamma = 2.0, \eta = 0.5$). Case ($i$) represents the deterministic evolution, where we do not monitor the dot, though some natural dephasing occurs (Sec.~\ref{sec:det}). Cases ($ii$), ($iii$) and ($iv$) involve continuous monitoring and implement the CUSUM-based technique (Sec.~\ref{sec:perform}). Case ($ii$) is the scenario of high monitoring efficiency (which, as we argue above, is likely unrealistic in practice), while in case ($iii$) the efficiency is low. Finally, in case ($iv$) the monitoring efficiency is moderate, but we have boosted the measurement strength to compensate. 

Physically, dephasing is intrinsic in any realistic system, as information about the state of the dot is lost to various internal modes, such as phonon modes.  In addition, there will be {\it additional} dephasing due to the presence of monitoring.  In these terms, case ($i$) has intrinsic dephasing but no monitoring; case ($ii$) has efficient monitoring and no intrinsic dephasing; case ($iii$) has monitoring with low efficiency, but no additional intrinsic dephasing. In case ($iv$) we assume that there is both monitoring and intrinsic dephasing, so the total dephasing rate $\gamma$ is higher than in case ($ii$), and the detector efficiency $\eta=0.5$ is moderate, reflecting the fact that only information from the monitoring is available, and not the information lost to the internal degrees of freedom.

Figure~\ref{fig:deffs} displays the single-photon probability as a function of $\Omega$ for these four cases. The parameter values are defined in Eq.~(\ref{eqn:parameters}), and we have set $g = 0.1$. This plot explores just a representative slice of the ($g$, $\Omega$) space that we considered earlier.  We see that case ($i$) performs consistently better than ($iii$), while ($ii$) unsurprisingly outperforms both ($i$) and ($iii$). However, case ($iv$) performs even better, and consistently dominates ($ii$). 

%
\section{Conclusions}
\label{sec:conclusions}
CUSUM is a simple yet powerful method that can significantly improve single-photon probability using continuous monitoring.  The protocol itself requires only simple components like an integrator and subtractor, along with $2$ registers and a comparator. This technique is useful in the weak coupling regime when the pumping rate is comparable to coupling strength.  In regions of strong coupling, or strong pumping, it is ineffective and can be worse than no monitoring at all.  We included various decoherence processes in our simulations, including spontaneous emission, dephasing and cavity leakage.  We modeled the electrical pumping and continuous monitoring  as a stochastic master equation. Numerical simulations showed that CUSUM performs quite well in regions of low $\Omega$, and significant improvements in single-photon probability were observed. We also studied the effect of imperfect monitoring efficiency on performance. 

\section*{ Acknowledgements}
The authors would like to thank Kaushik Choudhury, Dan Dapkus, Tony Levi, John O'Brien, Michelle Povinelli, and Alan Willner for useful discussions.  This work was supported in part by NSF Grant No.~ECS-0507270 and NSF CAREER Grant No.~CCF-0448658.

%


\begin{thebibliography}{10}

\bibitem{Knill01}
E. Knill, R. Laflamme and G.J. Milburn,  Nature {\bf 409}, 46 (2001).

\bibitem{Kok07}
P. Kok, W.J. Munro, K. Nemoto, T.C. Ralph, J.P. Dowling and G.J. Milburn, Rev. Mod. Phy. {\bf 79}, 135 (2007).

\bibitem{SPS-reviews}
M. Oxborrow and A. G. Sinclair,  Contem. Phy. {\bf 46}, 173 (2005); B. Lounis and M. Orrit, Rep. Prog. Phy. {\bf 68}, 1129 (2005).

\bibitem{SPS-QD-implementations}
J. Kim, O. Benson, H. Kan and Y. Yamamoto, Nature {\bf 397}, 500 (1999); C. Santori, D. Fattal, J. Vu\u{c}kovi\u{c}, G. S. Solomon and Y. Yamamoto, Nature {\bf 419}, 594 (2002); S. Laurent, S. Varoutsis, L. Le Gratiet, A. Lema"tre, I. Sagnes, F. Raineri, A. Levenson, I. Robert-Philip, and I. Abram, Appl. Phys. Lett. {\bf 87}, 163107 (2005); A.J. Shields, Nature Photonics {\bf 1}, 215 (2007).

\bibitem{Michler-Vahala03}
P. Michler (Ed.), \emph{Single Quantum Dots}, Springer (2003); K.J. Vahala, Nature {\bf 424}, 839 (2003); P. Michler (Ed.), \emph{Single Semiconductor Quantum Dots}, Springer (2009).

\bibitem{Raghunathan09}
S. Raghunathan and T.A. Brun, Phys. Rev. A {\bf 79}, 033831 (2009).

\bibitem{electrical-pumping}
A. Imamo$\breve{\text{g}}$lu and Y. Yamamoto, Phys. Rev. Lett. {\bf 72}, 210 (1994); P. Michler, A. Kiraz, C. Becher, W. V. Schoenfeld, P. M. Petroff, L. Zhang, E. Hu, and A. Imamoglu, $Science$ {\bf 290}, 2282 (2000); Z. Yuan,  B. E. Kardynal, R. M. Stevenson, A. J. Shields, C. J. Lobo, K. Cooper, N. S. Beattie, D. A. Ritchie, and M. Pepper, $Science$ {\bf 295}, 102 (2002); K. Sebald, P. Michler, T. Passow, D. Hommel, G. Bacher and A. Forchel, Appl. Phys. Lett. {\bf 81}, 2920 (2002); R. Hanson, L. P. Kouwenhoven, J. R. Petta, S. Tarucha and L. M. Vandersypen, Rev. Mod. Phys. {\bf 79}, 1217 (2007).

\bibitem{Levi07}
A.F.J. Levi, private communication.

\bibitem{SPS-theoretical}
A. Kiraz, M. Attat\"ure and A. Imamo$\breve{\text{g}}$lu, Phys. Rev. A {\bf 69}, 032305 (2004); F. Troiani, J.I. Perea, and C. Tejedor, Phys. Rev. B {\bf 73}, 035316 (2006); M.J. Fern$\acute{\mbox{e}}$e, H. Rubinsztein-Dunlop and G. J. Milburn, Phys. Rev. A {\bf 75}, 043815 (2007).

\bibitem{Trajectories}
T.A. Brun, Am. J. Phys. {\bf 70}, 719 (2002); K. Jacobs and D. A. Steck, Contemporary Physics {\bf 47}, 279 (2006).

\bibitem{Papoulis91}
A. Papoulis, \emph{ Probability, Random Variables and Stochastic Processes}, McGraw-Hill (1991); G. Grimmett and D. Stirzaker, \emph{Probability and Random Processes}, Oxford University Press (2006).

\bibitem{electrical-incoherent}
C.W. Gardiner and A. Eshmann, Phys. Rev. A {\bf 51}, 4982 (1995); O. Benson and Y. Yamamoto, Phys. Rev. A {\bf 59}, 4756 (1999).

\bibitem{stat-decision}
J.O. Berger, {\it Statistical Decision Theory And Bayesian Analysis}, Springer-Verlag (1985); P. H, Garthwaite, I.T. Jolliffe and B. Jones, {\it Statistical Inference}, Oxford University Press (2002); N. Mukhopadhyay and B.M. de Silva, {\it Sequential Methods and their Applications}, CRC Press (2009).

\bibitem{Press07}
W.H. Press, S.A. Teukolsky, W.T. Vetterling, and B.P. Flannery, {\it Numerical Recipes: The Art of Scientific Computing}, Cambridge University Press (2007).

\bibitem{Heindel10}
T. Heindel, C. Schneider, M. Lermer, S.H. Kwon, T. Braun, S. Reitzenstein, S. H$\ddot{\mbox{o}}$fling, M. Kamp, and A. Forchel, Appl. Phys. Lett. {\bf 96}, 011107 (2010). 
%
\end{thebibliography}
\end{document}